\documentclass[lettersize,journal]{IEEEtran}

\usepackage{amsmath,amsfonts}
\usepackage{algorithmic}
\usepackage{algorithm}

\usepackage{array}
\usepackage{float}
\usepackage{textcomp}
\usepackage{stfloats}
\usepackage{verbatim}
\usepackage{tabularx} 
\usepackage{txfonts} 
\usepackage{booktabs} 
\usepackage{multirow}

\usepackage{graphicx}
\usepackage{caption} 
\usepackage{subcaption} 
\usepackage{tcolorbox} 
\usepackage{tikz}

\usepackage{url}
\usepackage{cite}
\usepackage{hyperref}
\usepackage{cleveref}

\newenvironment{highlightbox}[1]{
    \begin{tcolorbox}[title={#1}]
    }{
    \end{tcolorbox}
}

\begin{document}

\newcommand \copyrighttext {
  \footnotesize \textcopyright 2024 IEEE. Personal use of this material is permitted. Permission from IEEE must be obtained for all other uses, in any current or future media, including reprinting/republishing this material for advertising or promotional purposes, creating new collective works, for resale or redistribution to servers or lists, or reuse of any copyrighted component of this work in other works. DOI: (to be added once issued) 
}

\newcommand\copyrightnotice{
    \begin{tikzpicture}[remember picture,overlay]
    \node[anchor=south,yshift=4pt] at (current page.south) {\fbox{\parbox{\dimexpr\textwidth-\fboxsep-\fboxrule\relax}{\copyrighttext}}};
    \end{tikzpicture}
}

\title{Replications, Revisions, and Reanalyses: Managing Variance Theories in Software Engineering}

\author{
    \IEEEauthorblockN{
        1\textsuperscript{st} Julian Frattini, 
        3\textsuperscript{rd} Davide Fucci,
        4\textsuperscript{th} Michael Unterkalmsteiner 
        5\textsuperscript{th} Daniel Mendez*} \\
    \IEEEauthorblockA{\textit{Blekinge Institute of Technology}
        Karlskrona, Sweden.
        \{firstname\}.\{lastname\}@bth.se} \\
    \and
    \IEEEauthorblockN{
        2\textsuperscript{nd} Jannik Fischbach}
    \IEEEauthorblockA{\textit{Netlight Consulting GmbH and *fortiss GmbH} Munich, Germany. jannik.fischbach@netlight.com}
    \thanks{Manuscript received December 2024}
}

\markboth{Transactions on Software Engineering,~Vol.~X, No.~Y, December~2024}%
{Shell \MakeLowercase{\textit{et al.}}: Managing Variance Theories in Software Engineering}


\maketitle

\copyrightnotice

\begin{abstract} 
Variance theories quantify the variance that one or more independent variables cause in a dependent variable.
In software engineering (SE), variance theories are used to quantify---among others---the impact of tools, techniques, and other treatments on software development outcomes.
To acquire variance theories, evidence from individual empirical studies needs to be synthesized to more generally valid conclusions.
However, research synthesis in SE is mostly limited to meta-analysis, which requires homogeneity of the synthesized studies to infer generalizable variance.
In this paper, we aim to extend the practice of research synthesis beyond meta-analysis.
To this end, we derive a conceptual framework for the evolution of variance theories and demonstrate its use by applying it to an active research field in SE.
The resulting framework allows researchers to put new evidence in a clear relation to an existing body of knowledge and systematically expand the scientific frontier of a studied phenomenon.
\end{abstract}


\section{Introduction}
\label{sec:intro}

Software engineering (SE) research aims to support SE practice in a process referred to as knowledge translation~\cite{santos2020research}.
It consists of knowledge \textit{creation}, which includes gathering empirical evidence in primary studies, and knowledge \textit{application}, i.e., providing this evidence to its target audience.
However, primary studies do not provide convincing decision support to practitioners on their own~\cite{lo2015practitioners,devanbu2016belief,franch2020practitioners,juristo2011role}.
Hence, an imperative step between the creation and application of knowledge is its \textit{synthesis}.
Research synthesis is a ``collective term for a family of methods that are used to summarize, integrate, combine, and compare the findings''~\cite{cruzes2011research} of individual pieces of evidence and aims to infer more generally valid conclusions.

One product of synthesizing quantitative research is a \textit{variance theory}.
Variance theories estimate the variance of a dependent variable in relation to one or more independent variables~\cite{ralph2018toward}.
In SE research, variance theories provide decision support by quantifying the strength of the effect of, for example, new tools, different technologies, or human factors on key performance indicators of the SE process.
For example, the synthesis of 27 primary studies about the effect of test-driven development (TDD) on code quality and developer productivity by Rafique and Mi{\v{s}}i{\'c} determined that ``TDD has a small positive effect on quality but little to no discernible effect on productivity''~\cite{rafique2012effects}.
However, research synthesis in SE is primarily limited to meta-analysis~\cite{santos2020research}.
While certain forms of meta-analysis excel at synthesizing evidence from quantitative studies, they only produce usable results under certain conditions like homogeneity of the pieces of evidence~\cite{hayes1999research}.
Current research synthesis practices fail to accommodate more complex relationships between individual pieces of evidence, like deviating hypotheses or the usage of different analysis methods.
Consequently, the validity of variance theories produced by these research synthesis practices is limited, and they may not offer the intended decision support to practitioners.

In this work, we propose a framework for managing variance theories in SE that extends beyond current meta-analysis practices.
We formally define quantitative, empirical evidence and an evolution framework that specifies how two pieces of evidence relate to each other.
We demonstrate the framework by applying it to an active field of SE research to show how it can guide SE research toward more coherent and productive research agendas.

We aim to support two scientific use cases.
First, our framework helps researchers to position new pieces of evidence in relation to an existing body of knowledge.
As such, the framework provides a terminology to frame how new studies advance the body of knowledge with respect to existing ones.
New evidence can be classified as either a replication, revision, or reanalysis, and the framework supports deciding whether this new evidence strengthens or challenges the body of knowledge.
Second, our detailed application demonstrates how to apply the framework to systematically review literature containing quantitative, empirical evidence.
By framing all empirical, quantitative studies investigating one phenomenon using the framework, the evolution of a variance theory about that phenomenon becomes tangible.
This reveals the current scientific frontier of quantitative studies on a phenomenon and can inform future study design.
Overall, our initiative aims to broaden the perspective on research synthesis to obtain more valid variance theories from SE research.

\textbf{Data Availability} All study data is publicly available in our replication package~\cite{tse2024replication}.

\section{Related Work}
\label{sec:related}

\subsection{Research Synthesis}
\label{sec:related:synthesis}

The purpose of any endeavor in SE research is to support SE practitioners~\cite{budgen2013case}.
This requires translating knowledge created in research to practice~\cite{beecham2014making,cartaxo2016evidence}.
However, previous research has shown that singular empirical studies do not provide convincing evidence to practitioners~\cite{lo2015practitioners}.
A single study cannot compete with the beliefs of practitioners, as the findings are limited to the context of the respective primary study~\cite{devanbu2016belief}.
Consequently, Miller advocated that the field of SE research ``needs to move to a portfolio of empirical studies (on a single research hypothesis) being the norm rather than the currently unconvincing ‘one-off’, normally laboratory-based, studies that currently dominate the research literature''~\cite{miller2005replicating}.
These portfolios of replications (also called \textit{families of studies or experiments}~\cite{basili1999building}) strengthen the validity of research findings and generate more reliable and general conclusions~\cite{nosek2020replication}.

While portfolios of replications strengthen the \textit{knowledge creation} phase of knowledge translation~\cite{santos2020research} by increasing the validity of drawn conclusions, they do not necessarily serve the \textit{knowledge application} phase.
Indeed, portfolios of replications complicate maintaining an overview of all relevant primary studies, and contradicting results are difficult to harmonize~\cite{rosenthal2001meta}.
This necessitates the aggregation and integration of results from primary studies~\cite{ciolkowski2005accumulation}, commonly referred to as \textit{research synthesis}.
Research synthesis describes ``methods that are used to summarize, integrate, combine, and compare the findings''~\cite{cruzes2011research} of primary studies with similar goals.
Shepperd summarized the synthesis process in five steps~\cite{shepperd2013combining}.

\begin{enumerate}
    \item \textbf{Problem formulation}: specifying a research question, usually about the influence of one independent on one dependent variable
    \item \textbf{Locating evidence}: searching literature that contains evidence about the research question
    \item \textbf{Appraising evidence quality}: applying quality inclusion criteria
    \item \textbf{Evidence synthesis and interpretation}: extracting relevant data and performing the research synthesis that infers about the initial research question based on the accumulated evidence
    \item \textbf{Reporting}: disseminating the results in a research report
\end{enumerate}

For step four, SE research has adopted several research synthesis methods from more mature experimental disciplines~\cite{dixon2005synthesising,pickard1998combining,shepperd2013combining,dieste2011comparative,santos2020research,santos2019procedure}.
Among the most popular is the narrative synthesis~\cite{santos2018analyzing}, a textual summary of research findings, often criticized for its lack of a systematic approach~\cite{ciolkowski2009we}.
The more systematic vote-counting classifies the effect of an independent variable on a dependent variable on ordinal scales, e.g., depending on its sign (i.e., positive, neutral, or negative effect)~\cite{pickard1998combining} or its strength of evidence (e.g., third party claims, circumstantial evidence, and strong evidence)~\cite{wohlin2013evidence}.
A histogram of classified findings from primary studies indicates the tendency of the effect observed by a portfolio of replications.

A more quantitative approach to synthesize results from controlled experiments is often called meta-analysis~\cite{rosenthal2001meta}.
Various forms of meta-analysis exist and are adopted in SE research~\cite{santos2018analyzing}.
The two approaches considered state-of-the-art are aggregate data (AD) meta-analysis and stratified independent participant data (IPD-S) meta-analysis~\cite{santos2018analyzing}.
AD meta-analysis pools the calculated effect sizes of primary studies together and calculates an overall effect size of the independent on the dependent variable~\cite{rosenthal2001meta}, often represented in a Forest plot~\cite{lewis2001forest}.
IPD-S meta-analysis pools together the raw data from all experiments of the primary studies and analyzes this data directly~\cite{riley2010meta}, but retains information about the belonging of each data point to the experiment it originated from to account for between-study variance~\cite{santos2018analyzing}.
IPD-S meta-analysis is commonly regarded as the gold standard for research synthesis~\cite{sutton2008recent} but is rarely applied in SE research~\cite{santos2018analyzing}.
Narrative synthesis and AD meta-analysis dominate SE research~\cite{santos2018analyzing}, though meta-analysis remains uncommon in general~\cite{kitchenham2020meta}.
Some examples include synthesizing evidence about defect prediction~\cite{hosseini2017systematic,zain2023application} and test-driven development~\cite{rafique2012effects}.

Research syntheses of quantitative evidence produce \textit{variance theories}, one of three commonly discussed types of theories~\cite{ralph2018toward}, about a phenomenon under study.
While \textit{theories for understanding} organize entities into meaningful categories and \textit{process theories} explain how something is happening~\cite{ralph2018toward}, variance theories quantify the strength of the effect of an independent on a dependent variable~\cite{ralph2018toward}.
Because they are a product of synthesis, variance theories have greater validity than a single piece of evidence~\cite{hannay2007systematic}.
The quantification of an effect strength offers decision support to practitioners, e.g., when deciding whether to adopt a technique like TDD.


\subsection{Shortcomings of the State-of-the-Art in Research Synthesis}
\label{sec:related:shortcomings}


Literature has acknowledged several shortcomings of the state-of-the-art of research synthesis in SE~\cite{shepperd2013combining,cruzes2011research,kitchenham2020meta}.

\paragraph{Dealing with heterogeneity}
Primary studies involved in research synthesis are subject to heterogeneity, i.e., differences between studies investigating the same phenomenon~\cite{ciolkowski2009we}.
An unclear study design, incomplete sample selection protocol, or unreflected operationalization of latent concepts in SE research often obscure critical factors~\cite{shepperd2013combining,pickard1998combining} like prior knowledge of participants, their experience, or intrinsic motivation.
If these factors have a significant influence on the phenomenon under investigation, traditional meta-analysis techniques will produce inconclusive results.

When the factors causing these differences are well understood and recorded in the data collection process, an appropriate synthesis method can account for them~\cite{hayes1999research,pickard1998combining}, which benefits the validity of the conclusion~\cite{nosek2020replication}.
However, traditional meta-analysis techniques are limited to studies investigating the relationship between exactly two variables (i.e., the effect of one independent on one dependent variable)~\cite{miller2000applying,santos2018analyzing}.
Yet, phenomena in SE research can rarely be isolated into a two-variable relationship, as human~\cite{pickard1998combining} and other context factors~\cite{pfleeger1999albert} usually interact with the phenomenon.
If the body of primary studies is very large, these context factors may be analyzed by conducting a meta-analysis of meta-analyses, as shown by Harris et al. in a study with 136 primary studies clustered into 31 meta-analyses~\cite{harris1985mediation}.
Similarly, Rafique and Mi{\v{s}}i{\'c} conducted a meta-analysis of 27 primary studies about the effect of TDD on developer productivity and code quality, where the amount of evidence allowed a subgroup analysis~\cite{rafique2012effects}.
This subgroup analysis revealed that the population from which study participants were drawn (practitioners vs. students) mediated some of the observed effects.
However, meta-meta-analyses or subgroup analyses are no reliable tool to address the presented shortcoming for the following reasons.
Firstly, the required amount of empirical evidence is mostly unavailable in SE research~\cite{basili1999building}.
Secondly, such analyses are only eligible if the synthesized primary studies report additional information, such as the sample demographics.
Finally, these analyses only add a hierarchical complexity to the research synthesis endeavor but do not systematically deal with more complex relationships between variables~\cite{sutton2001bayesian}.
Consequently, classical research synthesis methods like meta-analysis only apply to a set of homogeneous primary studies.
They fail to incorporate that causal assumptions may evolve and become more complex than simple two-variable relationships.

\paragraph{Limited to experimental studies}
Additionally, many synthesis methods are limited in the types of primary studies they can integrate.
Traditional meta-analyses from medical research are constrained to controlled experiments~\cite{rosenthal2001meta}.
This excludes, by design, quantitative evidence from observational studies~\cite{sutton2001bayesian} and all qualitative studies~\cite{roberts2002factors}.
However, this discards evidence about complex phenomena that might not be studied using controlled experiments alone~\cite{roberts2002factors}.
On the one side, observational studies are less invasive to the actual software development context but can still yield reliable, causal inferences using appropriate methods~\cite{mcelreath2018statistical,furia2022applying}.
Conversely, qualitative studies have shown to capture richer context information, especially in medical and social sciences~\cite{dixon2005synthesising}, which has also been acknowledged in SE research~\cite{cruzes2011research}.


\paragraph{Static and retrospective}
Finally, research syntheses receive critique for being static when they ``should be updated on completion of a study to place their result in context''~\cite{sutton2008recent}.
A continuous approach to meta-analyses~\cite{ciolkowski2009we} would avoid that included studies become outdated by the time of synthesis~\cite{jaccheri2021systematizing}.
Furthermore, research synthesis is predominantly conducted retrospectively, i.e., it aggregates publications published prior but rarely guides the design of future ones~\cite{dyba2007applying}.
Instead of opportunistically conducted meta-analysis~\cite{santos2018analyzing}, SE research requires prospective synthesis initiatives as also called for in other disciplines like medical research~\cite{riley2010meta}.

\section{Goal and Method}
\label{sec:method}

We aim to address the shortcomings outlined in~\Cref{sec:related:shortcomings} with a strategy to facilitate more coherent, systematic research.
To this end, we composed a framework from the scientific practices of three branches of research evolution.
First, we surveyed the existing body of knowledge on replication studies in SE~\cite{baldassarre2014replication,gomez2010replications} and their synthesis in the form of meta-analysis~\cite{hayes1999research,dieste2011comparative,shepperd2013combining}.
Second, we reviewed techniques from statistical causal inference~\cite{siebert2023applications}, particularly model comparison for causal inference from observational studies~\cite{furia2022applying,pearl2010causal,cinelli2024crash}.
Finally, we surveyed references about the evolution of statistical practices~\cite{torkar2020bayesian,furia2019bayesian,mcelreath2018statistical}

A regular validation of the framework would require applying it in practice, i.e., comparing the evolution of variance theories with and without the framework.
However, this kind of validation is not feasible at this point as it would require its prior adoption.
Instead of developing the framework, applying it in a controlled manner, and validating it with collected data, we sought to involve the SE research community at an early stage of proposing this framework already to allow for critical discussions and contributions.
Hence, we opted for a constructive validation of the proposed framework in a focus group setup.
We presented the framework at the annual meeting of the International Software Engineering Research Network (ISERN).
The ISERN community\footnote{\url{https://isern.iese.de/}} consists of experts on empirical software engineering methodologies and their applications and meets as part of the Empirical Software Engineering International Week.\footnote{\url{https://conf.researchr.org/series/esem}}
In our focus group session, we discussed the eligibility of the framework to guide future empirical research and allow effective aggregation of evidence.
Based on the discussion, we revised the framework and the boundary conditions of its applicability.

\section{Conceptual Framework}
\label{sec:framework}

Our framework for managing variance theories consists of a definition of evidence in \Cref{sec:framework:evidence} and a flowchart describing the evolution of evidence in \Cref{sec:framework:evolution}.
\Cref{sec:framework:impact} describes the implications of the framework on research synthesis practices in SE.

\subsection{Evidence}
\label{sec:framework:evidence}

We define a piece of empirical, quantitative evidence $e$ as a tuple $e:=E(h, d, m)$ consisting of three components.

\begin{itemize}
    \item \textbf{Hypothesis} $h$: A hypothesis consisting of variables and (assumed) causal relationships between those variables. For example, $h_1:=x \rightarrow y$ defines hypothesis $h_1$ as variable $x$ causally influencing variable $y$.
    \item \textbf{Data} $d$: A record of observations of all variables contained in $h$. An eligible dataset $d_1$ for $h_1$ requires observation for both variables $x$ and $y$.
    \item \textbf{Method} $m$: An analysis method that processes the data $d$ under the hypothesis $h$ to produce a conclusion.
\end{itemize}

\begin{figure}
    \centering
    \begin{subfigure}[b]{0.2\textwidth}
         \centering
         \includegraphics[width=\textwidth]{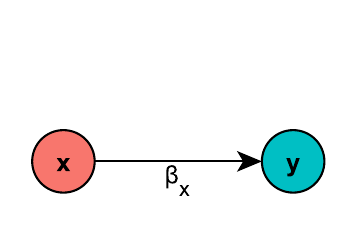}
         \caption{Basic hypothesis $h_1$}
         \label{fig:dag:basis}
     \end{subfigure}
     \hfill
     \begin{subfigure}[b]{0.2\textwidth}
         \centering
         \includegraphics[width=\textwidth]{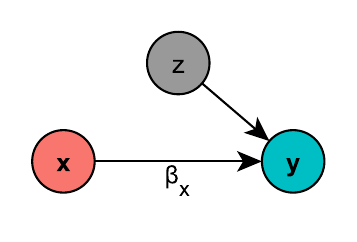}
         \caption{Revision $h_2$ increasing the effect estimate precision}
         \label{fig:dag:precision}
     \end{subfigure}
     \hfill
     \begin{subfigure}[b]{0.2\textwidth}
         \centering
         \includegraphics[width=\textwidth]{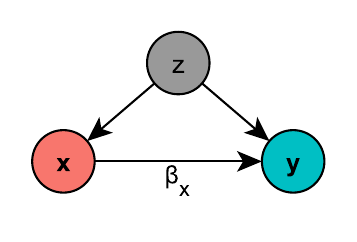}
         \caption{Revision $h_3$ to deconfound the effect estimate}
         \label{fig:dag:confounder}
     \end{subfigure}
     \hfill
     \begin{subfigure}[b]{0.2\textwidth}
         \centering
         \includegraphics[width=\textwidth]{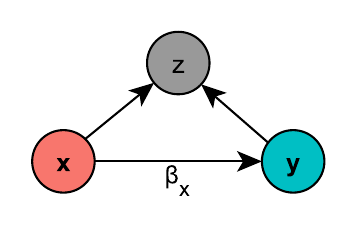}
         \caption{Revision $h_4$ specifying a collider}
         \label{fig:dag:collider}
     \end{subfigure}
    \caption{DAGs representing a hypothesis and three revisions}
    \label{fig:dags}
\end{figure}

Hypotheses are networks of variables and relationships among them.
As such, they can be visualized via directed, acyclic graphs (DAGs), as shown in \Cref{fig:dags}.
In these DAGs, nodes represent variables, and directed edges represent assumed causal relationships.
\Cref{fig:dag:basis} shows a graphical representation of a simple, two-variable hypothesis.
More often, though, manuscripts present such simple hypotheses textually.
This often takes the form of a verbose null hypothesis, e.g., ``There is no significant difference in values of $y$ for different values of $x$.''
More complex hypotheses involving more variables and relationships like \Cref{fig:dag:precision,fig:dag:confounder} require graphical representation but are rare in SE research~\cite{siebert2023applications}.

The two colored nodes in the four DAGs represent the main \textit{phenomenon} of interest, i.e., the independent variable (colored red) and the outcome or response variable (colored cyan).
The goal of a piece of empirical, quantitative evidence is to estimate the average causal effect (ACE) of the main independent variable(s) on the dependent outcome variable.
Additional variables (colored grey) may be relevant to the hypothesis but not part of the main phenomenon under study.
Therefore, the same phenomenon of interest can be involved in multiple hypotheses.

All analysis methods $m$ require deriving a \textit{statistical model} from the causal model, i.e., the hypothesis $h$~\cite{mcelreath2018statistical}.
A statistical model typically consists of a regression model, i.e., a specified, often linear relationship between one or more predictors and the outcome variable.
In the case of simple, two-variable hypotheses, this boils down to regressing the outcome on the only predictor.
For example, the statistical model derived from $h_1$ in \Cref{fig:dag:basis} would be $y \sim x$.
In the case of more complex hypotheses, one must select the subset of independent variables.
This subset should maximize the precision of the estimation of the effect of $x$ on $y$ and, on the other hand, ensure that the causal effect is not confounded.
For example, the statistical model derived from $h_3$ shown in \Cref{fig:dag:confounder} would be $y ~ x + z$ where $z$ is included to de-confound the effect of $x$ on $y$.
The statistical model of $h_4$ shown in \Cref{fig:dag:collider} would be $y ~ x$ since including $z$, called a \textit{collider}, would confound $x$ on $y$~\cite{mcelreath2018statistical}. 
The subset that de-confounds the causal effect is commonly called the \textit{adjustment set}~\cite{cinelli2024crash} and can be determined by applying a systematic procedure called the \textit{backdoor adjustment}~\cite{pearl2010causal}.
Explaining this procedure in detail goes beyond the scope of this manuscript but is well-explained in existing literature~\cite{mcelreath2018statistical,cinelli2024crash,pearl2010causal}.

The eligibility of analysis methods depends on the complexity of the hypothesis and the properties of the variables.
For simple, two-variable hypotheses consisting of one independent and one dependent variable, most scholars resort to null hypothesis significance tests (NHSTs) like the Student's t-test or its variants.
Here, the choice depends on the normality of the dependent variable and whether the data is paired or not.
For more complex hypotheses involving more than one independent variable, scholars tend to apply linear regression models with multiple predictors.

Applying the analysis method $m$ to the data set $d$ based on the causal model implied by the hypothesis $h$ produces the piece of evidence $e$ that offers a conclusion.
The nature of the conclusion depends on the analysis method.
For example, NHSTs propose a p-value that scholars commonly compare with an arbitrary significance level $\alpha$ to determine whether the independent variable evokes a statistically significant difference in the dependent variable.
For linear regression models, the conclusion takes the form of coefficients (e.g., $\beta_x$ in \Cref{fig:dag:basis}, representing the strength of the impact of $x$ on $y$).
From these coefficients, one can additionally calculate confidence intervals for each independent variable.
If the confidence interval of a variable is not consistent with 0, i.e., it does not intersect 0, then the variable is considered to have a significant impact on the dependent variable.

\subsection{Evolution of Evidence}
\label{sec:framework:evolution}

\begin{figure*}
    \centering
    \includegraphics[width=\linewidth]{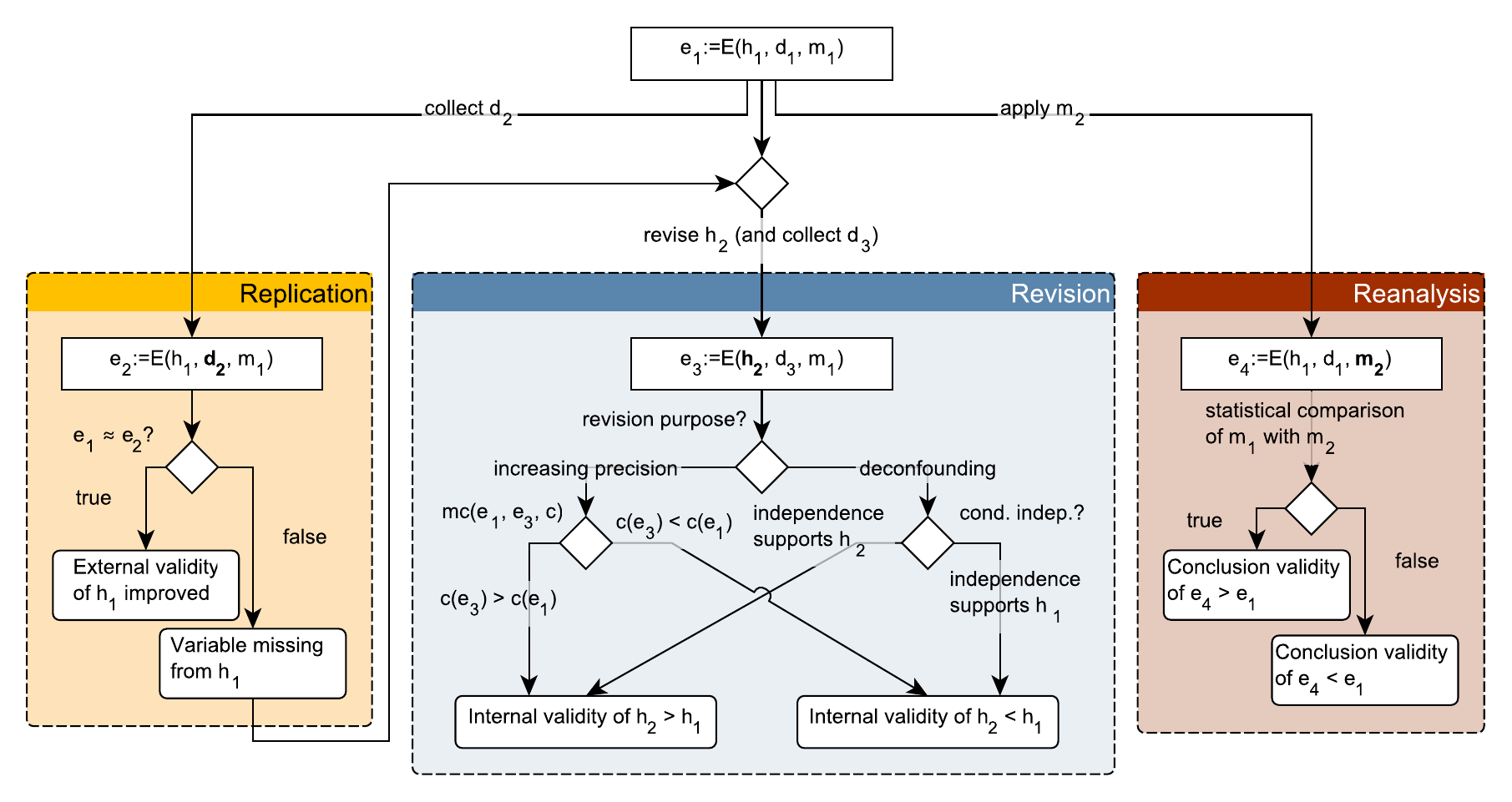}
    \caption{Framework describing the evolution of quantitative, empirical evidence}
    \label{fig:framework}
\end{figure*}

Variance theories emerge from the synthesis of multiple pieces of empirical, quantitative evidence, which increases their validity and abstracts from passing trends~\cite{hannay2007systematic}.
To accommodate research synthesis that goes beyond the meta-analysis of a homogeneous set of primary studies~\cite{pickard1998combining}, the relationship between two pieces of evidence needs to be clear.
\Cref{fig:framework} visualizes the types of evolution of empirical evidence, i.e., the possible relationship between pieces of evidence.
Starting from an initial piece of evidence $e_1=E(h_1, d_1, m_1)$, we consider three types visualized as paths in \Cref{fig:framework} and explained next.

\subsubsection{Replication}
\label{sec:framework:evolution:replication}

The most commonly known evolution type of empirical evidence in SE research is through \textit{replication} (left branch colored yellow in \Cref{fig:framework}).
A replication is a type of study that offers diagnostic evidence about a previous empirical study~\cite{nosek2020replication}.
As such, a replication subscribes to the same causal hypothesis $h_1$ and uses the same analysis method $m_1$ but collects a different data set $d_2$ to produce a new piece of evidence $e_2:=E(h_1, d_2, m_1)$.
The conclusion derived from the replication $e_2$ is compared with the conclusion of the original piece of evidence $e_1$ to check for agreement.
Checking for agreement depends on the nature of the conclusion that the analysis method $m_1$ produces.
If $m_1$ is a type of hypothesis test that produces a p-value, this check is referred to as an \textit{aggregation of p-values}~\cite{santos2018analyzing} via Fisher's or Stouffer's method~\cite{borenstein2021introduction}.
If $m_1$ is a type of regression model that produces confidence intervals of coefficients, then the check boils down to assessing whether the confidence intervals overlap in a Forest plot~\cite{lewis2001forest}, or AD or IPD-S meta-analysis techniques~\cite{santos2018analyzing}.
If the conclusions agree, the external validity of the causal claim $h_1$ is improved as the replication shows that the conclusion of $e_1$ also holds in a different context $d_2$.
However, in case the conclusions disagree, SE literature offers little advice on how to relate these results.
The disagreeing conclusions indicate that at least one variable that would explain the difference between $e_1$ and $e_2$ is missing from $h_1$, requiring a \textit{revision} of the original hypothesis.

\subsubsection{Revision}
\label{sec:framework:evolution:revision}

A less commonly discussed evolution of empirical evidence in SE research is through \textit{revision} of a hypothesis (middle branch colored blue in \Cref{fig:framework}).
Revising hypothesis $h_1$ means proposing a competing network of variables, hypothesis $h_2$, that supposedly explains the phenomenon under study---which produced the data $d_1$ and $d_2$---better and, therefore, has greater internal validity.
The competing network can include new or discard existing variables or may---alternatively or additionally---propose different causal relationships between variables.
Only the variables belonging to the phenomenon under study need to remain included.
Otherwise, the new hypothesis pertains to a different phenomenon.
\Cref{fig:dag:precision,fig:dag:confounder,fig:dag:collider} visualize revisions of \Cref{fig:dag:basis} as they contain the variables of the main phenomenon under study ($x$ and $y$) but include an additional variable ($z$) and different relationships.

Revisions can serve two different purposes~\cite{cinelli2024crash}.
The first purpose is to increase the precision of estimating the effect $\beta_x$ of $x$ on $y$.
For example, involving an additional, independent variable $z$ with an assumed causal relationship $z \rightarrow y$ may increase the precision of the estimate of the average causal effect~\cite{cinelli2024crash}.
\Cref{fig:dag:precision} visualizes such a revised hypothesis $h_2$.
The second possible purpose of a revision is to de-confound the estimation of the effect $\beta_x$ of $x$ on $y$~\cite{mcelreath2018statistical}.
This is particularly relevant to phenomena studied in observational, not experimental, settings where the independent variable of interest can be influenced by factors other than random assignment.
A confounder could be a common cause as visualized in \Cref{fig:dag:confounder} where variable $z$ impacts both $x$ and $y$, therefore biasing the direct effect of $x$ on $y$.
The adjustment set of this hypothesis includes $z$ as a predictor of $y$ to de-confound the effect of $x$ on $y$~\cite{mcelreath2018statistical}.

The disagreeing conclusions from a replication but also emerging qualitative evidence (represented by the direct arrow from $e_1$ to $e_3$ in \Cref{fig:framework}) may trigger a revision.
For example, a qualitative study might suggest that the variable $z$ also influences $y$ even before observing disagreeing conclusions from replications.
Proposing a new hypothesis $h_2$ may also require collecting a new data set $d_3$ if $h_2$ contains variables not recorded in $d_1$.
In the abstract example, a new data set $d_3$ that records both $z$ in addition to $x$ and $y$ is necessary. 

Once appropriate data are available, the competing hypotheses are evaluated by model comparison.
The type of comparison depends on the purpose of the revision.
To evaluate a revision aiming at increasing the precision, the out-of-sample predictive power of the two models is compared (abbreviated as \texttt{mc} in \Cref{fig:framework}) via an appropriate criterion (abbr. as \texttt{c}).
Metrics like the Akaike Information Criterion (AIC)~\cite{akaike2011akaike} or leave-one-out cross-validation (LOO) may be applied depending on the analysis method~\cite{vehtari2017practical,magnusson2020leave}.
These metrics assign scores to competing hypotheses and infer which predicts the observed data best.
The model with the greatest predictive power is assumed to be more internally valid.

To evaluate a revision aiming at de-confounding, testable implications in the form of \textit{independencies} and \textit{conditional independencies} are derived from the hypotheses~\cite{mcelreath2018statistical}.
For example, according to $h_3$ in \Cref{fig:dag:confounder}, both $x$ and $y$ depend on $z$. 
Conversely, $h_1$ in \Cref{fig:dag:basis} does not include this claim and implies that $x$ and $y$ are independent of $z$. 
Additionally, $h_3$ implies that the strength of the ACE of $x$ on $y$ changes when conditioning on $z$ via deconfounding, which $h_1$ does not imply.
Correlational analyses on the available data set $d_3$ can confirm or refute these assumed independencies~\cite{mcelreath2018statistical}.
Comparing the statistical model $y \sim x$ (derived from $h_1$) with $y \sim x + z$ (derived from $h_3$) produces two estimates of the ACE of $x$ on $y$.
If the ACE is the same, then the effect of $x$ on $y$ is independent of $z$ and $h_1$ represents the causal relations of the phenomenon under study better.
If the ACE is different, then the effect depends on $z$, and $h_3$ is more valid.
Consequently, the internal validity of $h_3$ exceeds the one of $h_1$ and can be considered the currently superior causal model to explain the phenomenon under investigation.

While both purposes of revisions aim to strengthen the internal validity of a hypothesis, the respective evaluation that decides the comparison is not interchangeable.
Model comparison used to determine the hypothesis with the greater out-of-sample predictive power is not fit when aiming to deconfound a hypothesis, as a confounded hypothesis may very well exhibit a greater predictive power than a deconfounded one~\cite{mcelreath2018statistical}.
Consequently, the distinction of purpose when conducting a revision is imperative for the choice of evaluation method.

The hypothesis that currently shows the greatest internal validity should be the one that future studies should subscribe to.
This means that all future studies investigating the phenomenon of $x$ and $y$ should record all variables involved in the hypothesis that are part of the adjustment set.

\subsubsection{Reanalysis}
\label{sec:framework:evolution:reanalysis}

The least commonly discussed evolution of empirical evidence in SE research is the \textit{reanalysis} of existing data (right branch colored red in \Cref{fig:framework}).
Reanalysis---sometimes also referred to as a ``test for robustness''~\cite{nosek2020replication}---describes the application of a different analysis method $m_2$ to the same data $d$ under the same causal assumptions $h$~\cite{gomez2010replications}.
In special cases, however, reanalysis may also be necessitated by a revision.
For example, extending a hypothesis to include two instead of one independent variable will make analysis methods that only operate with one independent variable (e.g., a t-test) ineligible and necessitate more complex ones (e.g., a linear model).

Reanalyses are mostly driven by adapting more advanced methods from other disciplines (e.g., statistics, or medical research).
One instance of this type of evolution is the ongoing endeavor to abandon simple NHSTs for more advanced Bayesian data analysis~\cite{furia2019bayesian}.
Reanalyses increase the conclusion validity of the evidence by revising statistical assumptions~\cite{wohlin2012experimentation}.
The decision of which analysis method to prefer over another is often based on intricate statistical comparisons~\cite{mcelreath2018statistical}, which SE researchers usually adapt and do not conduct themselves.

\subsection{Implications on Research Synthesis}
\label{sec:framework:impact}

The framework for systematic evolution of variance theories has the following properties that address two of the three shortcomings in research synthesis described in \Cref{sec:related:shortcomings}.

\begin{enumerate}
    \item \textbf{Causal}: The framework takes a causal perspective to research synthesis by adding \textit{revisions} as a type of evolution next to \textit{replications}.
    This allows systematically addressing the heterogeneity of evidence conclusions by evolving hypotheses beyond two-variable relations.
    \item \textbf{Open to different study types}: By abandoning the constraint of simple, two-variable relations, the framework for research synthesis also opens up to other study types.
    More complex networks of variables can adequately represent causal assumptions from an observational study~\cite{furia2022applying}.
    Additionally, the framework allows qualitative studies to contribute to the evolution of quantitative variance theories by triggering revisions.
\end{enumerate}

How to address the remaining, third shortcoming in research synthesis will be discussed in \Cref{sec:discussion:future}.

\section{Application}
\label{sec:application}

We apply our framework to the research field of requirements quality.
\Cref{sec:application:background} introduces the relevant background about the research field.
\Cref{sec:application:evolution} demonstrates the application of the framework to a set of primary studies from this field.
Finally, \Cref{sec:application:critique} retraces omitted steps between existing pieces of evidence to show how the framework aids in systematically evolving evidence.
For readability, \Cref{sec:application:evolution,sec:application:critique} omit details on statistical operations to focus on the evolution on a conceptual level.
Details on the statistical operations can be found in our replication package~\cite{tse2024replication}.

\subsection{Requirements Quality Research}
\label{sec:application:background}

Requirements quality research~\cite{montgomery2022empirical} is concerned with identifying how properties of requirements artifacts~\cite{frattini2022live} (e.g., passive voice, sentence length) impact properties of subsequent software development activities (e.g., correctness of implementing, completeness of testing) that use these requirements artifacts as part of their input~\cite{frattini2024measuring}.
Variance theories in this field quantify the strength of the effect that certain properties of requirements artifacts have---e.g., whether longer requirements sentences reduce the correctness when implementing source code~\cite{ferrari2018detecting}.
Variance theories inform requirements writing guidelines by indicating to practitioners whether addressing these properties is worth it~\cite{frattini2023requirements}---e.g., whether reducing the sentence length of requirements should be enforced.
One such property of requirements artifacts commonly discussed in requirements quality literature is the use of \textit{passive voice} in natural language (NL) requirements.
The following two versions of the same requirement illustrate the difference.

\begin{itemize}
    \item \textbf{Active}: The system shall obtain all transaction details from the Statement Database.
    \item \textbf{Passive}: All transaction details shall \textit{be obtained} from the Statement Database.
\end{itemize}

The passive formulation omits the actor (``the system'') from the requirements specification, obscuring who is allowed to perform the action.
The observed drop in informativeness caused textbooks to advise against using passive voice in NL requirements artifacts~\cite{pohl2016requirements}.
However, the lack of empirical evidence for this claim, and even evidence against it~\cite{krisch2015myth} attracted experimental studies investigating its impact.

\subsection{Evolution of Evidence}
\label{sec:application:evolution}

We are aware of three studies that empirically investigate the impact that the use of passive voice in NL requirements has on the domain modeling activity~\cite{femmer2014impact,frattini2024second,frattini2024applying}. 
The three studies arrive at the variance theory that passive voice has a slight negative impact on the completeness of domain models.

These three studies contribute four pieces of evidence as one of the studies produces two separate pieces of evidence~\cite{frattini2024applying}.
\Cref{tab:application:quality} lists these four pieces of evidence indexed as $e_1$-$e_4$.
The figure in the leftmost column of the table visualizes the relationship between the pieces of evidence as a version control graph (VCG).
The color of each node in this graph reflects the evolution type that this piece of evidence represents in relation to its predecessor based on the colors in \Cref{fig:framework}.
For example, $e_3$ is a replication of $e_1$, hence the box is yellow.

\begin{table*}[!ht]
    \centering
    \caption{Evolution of the variance theory on the impact of passive voice on domain modeling}
    \label{tab:application:quality}
    \begin{tabular}[b]{lllllll}
        \toprule
        \textbf{VCG} & \textbf{Evolution Type} & \textbf{Ref.} & \textbf{Hypothesis} & \textbf{Data} & \textbf{Analysis Method} & \textbf{Conclusion} \\
        \midrule
        \multirow[c]{4}{*}{ \includegraphics[height=34px]{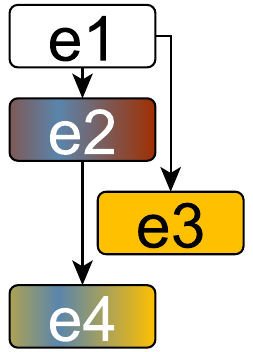}} & Original Study & 
        \cite{femmer2014impact} & $h_1$ & $d_1$ & $m_1$: Mann-Whitney U test & $p=0.001$ \\
        & Revision \& Reanalysis & \cite{frattini2024second} &  $h_2$ & $d_1$ & $m_2$: Bayesian model & $[-0.17, \sim0.49, +0.34]$ \\
        & Replication & \cite{frattini2024applying} &  $h_1$ & $d_2$ & $m_1$: Wilcoxon signed-rank test & $p=0.025$ \\
        & Revision \& Replication & \cite{frattini2024applying} & $h_3$ & $d_2$ & $m_2$: Bayesian model & $[-0.25, \sim0.30, +0.45]$ \\
        \bottomrule
    \end{tabular}
\end{table*}

\subsubsection{Original study}
\label{sec:application:evolution:esem}

\begin{figure}
    \centering
    \includegraphics[width=\linewidth]{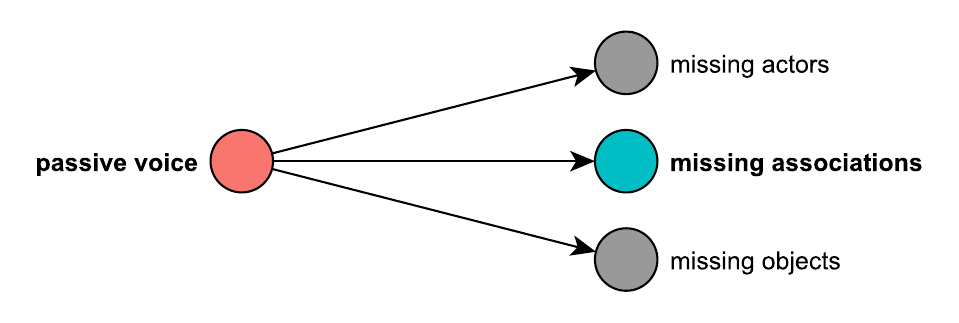}
    \caption{Hypothesis $h_1$ investigated by Femmer et al.~\cite{femmer2014impact}}
    \label{fig:rq:h1}
\end{figure}

To the best of our knowledge, Femmer et al. contributed the first piece of empirical, quantitative evidence $e_1$ about the impact of passive voice on domain modeling.
In their study~\cite{femmer2014impact}, they investigated the research question ``Is the use of passive sentences in requirements harmful for domain modelling?''
In particular, they studied whether the use of passive voice in NL requirements sentences changed the number of missing actors ($Act^-$), associations ($Asc^-$), and domain objects ($Obj^-$) from domain models derived from them.
\Cref{fig:rq:h1} visualizes their three hypotheses in one DAG.
In this demonstration, we focus on the impact of passive voice on the number of missing associations, i.e., $h_1: passive \rightarrow Asc^-$.

Femmer et al. conducted a parallel-design controlled experiment with 15 university students as participants.
These participants were randomly divided into either the control or the treatment group.
Each participant received seven natural language requirements sentences that were either written in active voice (for the control group) or passive voice (for the treatment group).
The experimental task was to generate a domain model from each requirement.
Then, the authors of the study counted the number of missing actors, associations, and domain objects from the resulting domain models via comparison to a gold standard that they produced.
The resulting data set $d_1$ consequently consisted of 105 ($=15 \times 7$) data points recording the group (active or passive), the number of missing actors, associations, and domain objects, as well as several demographic factors like program and experience.

To produce evidence $e_1$, the authors applied a Mann-Whitney U test to evaluate the three hypotheses.\footnote{In the original study~\cite{femmer2014impact}, the authors use the per-participant aggregate of missing entities as a response variable, which we do not to stay true to the hypothesis. The conclusion remains the same.}
The conclusion of evidence $e_1=E(h_1, d_1, m_1)$ is $p=0.001$, i.e., the NHST suggests that the use of passive voice has a statistically significant impact on the number of missing associations from a domain model.
The authors report an effect size calculated via Cliff's $\delta$ of 0.75~\cite{femmer2014impact}, which suggests a strong effect on per-participant aggregate level.
Our re-calculation on domain model level amounts to an effect size of only 0.35, which is considered of medium strength.

\subsubsection{Follow-up Study 1}
\label{sec:application:evolution:wsese}

\begin{figure}
    \centering
    \includegraphics[width=\linewidth]{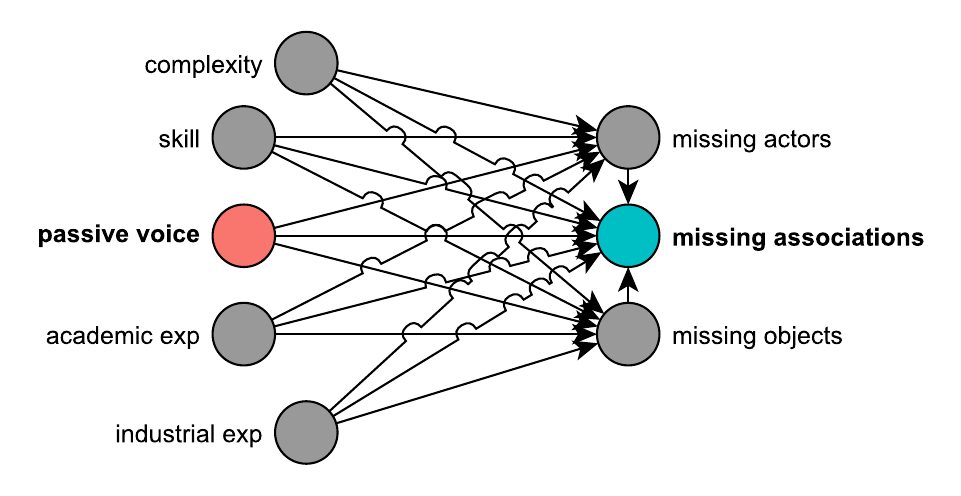}
    \caption{Hypothesis $h_2$ revised by Frattini et al.~\cite{frattini2024second}}
    \label{fig:rq:h2}
\end{figure}

Frattini et al. performed a follow-up study, taking a second look at the drawn conclusions~\cite{frattini2024second} to produce evidence $e_2$.
In this study, they reused the collected data $d_1$, but both propose a new causal hypothesis $h_2$ and also applied a different analysis method $m_2$.
The revised causal hypothesis $h_2$ contained the following additional assumptions which are also visualized in \Cref{fig:rq:h2}:

\begin{enumerate}
    \item If either actors or domain objects are missing, associations are more likely to be missing as well as one of the nodes involved in the edge is not present.
    \item The general skill of a participant may influence their performance.
    \item The complexity of a requirement may affect how easy it is to miss an actor, association, or domain object.
    \item The academic and industrial experience of a participant may influence their performance.
\end{enumerate}

The first additional assumption added two relationships to the DAG, and the second to fourth added new variables.
Data set $d_1$ already recorded values for these additional variables.
In a survey prior to the experiment, participants reported their academic and industrial experience on an ordinal scale with four categories (i.e., no experience, up to 6 months, 6 to 12 months, and more than 12 months).

The number of predictors involved in a statistical model derived from the causal model $h_2$ made the analysis method $m_1$ ineligible for producing a conclusion, as the Mann-Whitney U test only operates with one predictor but the statistical model requires several.
Instead, the authors followed the advice to adopt Bayesian data analysis~\cite{furia2019bayesian,mcelreath2018statistical} as their method $m_2$.
The resulting statistical model used the available demographic factors as predictors and models the requirements' complexity and participants' skill as random effects via the IDs of the requirements and participants.

Applying a Bayesian data analysis $m_2$ under the causal assumptions encoded in $h_2$ to the existing data $d_1$ produced a marginal probability distribution of the impact of passive voice on the number of missing associations.
Evidence $e_2=E(h_2, d_1, m_2)$ concludes that the use of passive voice in NL requirements leads to more missing associations in about 34\%, to fewer in 17\%, and to an equal amount in 48\% of all cases on average.
Hence, $e_2$ agrees with $e_1$ regarding effect strength~\cite{femmer2014impact}, but the Bayesian data analysis $m_2$ cautions about the significance of the effect suggested by $m_1$.
The authors of the follow-up study referred to literature for the superiority of $m_2$ over $m_1$ but did not validate whether $h_2$ was more valid than $h_1$.

\subsubsection{Follow-up Study 2}
\label{sec:application:evolution:emse}

Frattini et al. performed a second follow-up study where they conducted their own experiment~\cite{frattini2024applying}.
In this crossover-design experiment involving 25 participants, mostly from industry, the experimental task was similar, but the material differed (i.e., it used four different NL requirements).
Additionally, due to the crossover design, participants were not divided into a treatment and control group but received all levels of the treatment (just in different orders). 
Also, the experiment involved another treatment (the use of ambiguous pronouns), which represents a different phenomenon and, hence, is irrelevant in this context.
The resulting data set $d_2$ consisted of 100 ($=25 \times 4$) data points.

This second follow-up study~\cite{frattini2024applying} produced two pieces of evidence.
First, the authors performed a replication of $e_1$ by applying the same analysis method $m_1$ under the same causal assumptions $h_1$ to the new data $d_2$.
Since the data is paired due to the crossover design, the configuration of $m_1$ changed slightly (making it necessary to use a Wilcoxon signed-rank test instead of a Mann-Whitney U test), but the inferential process remains comparable.
Evidence $e_3=E(h_1, d_2, m_1)$ concludes that passive voice has a statistically significant impact on the number of missing associations with $p=0.025$.
Although the p-values of $e_1$ and $e_3$ differ, both reject the null hypothesis of no impact under the common level of significance $\alpha=0.05$ and, therefore, suggest the same conclusion.

\begin{figure}
    \centering
    \includegraphics[width=\linewidth]{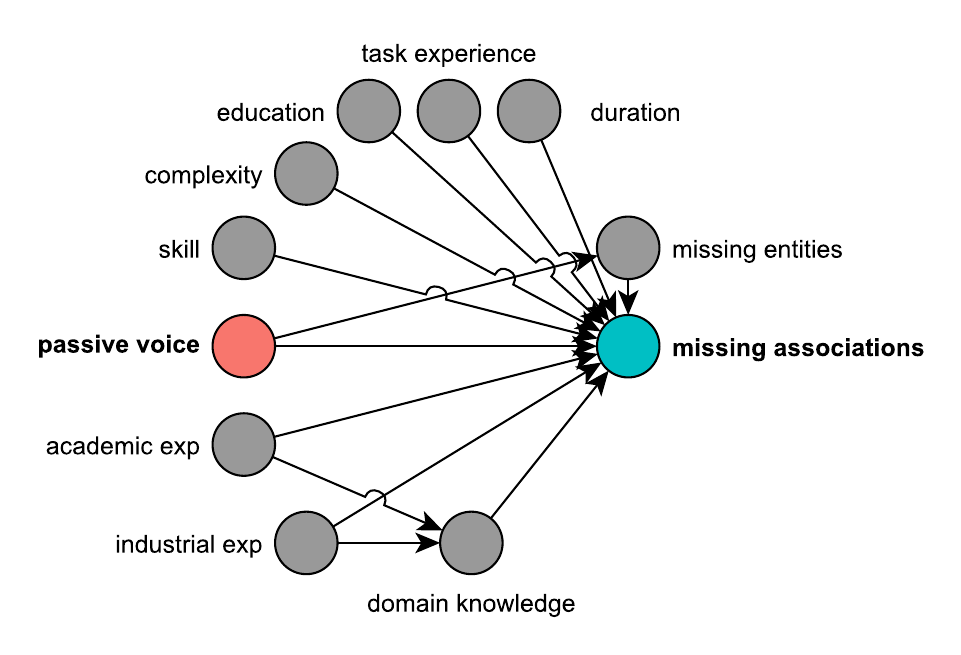}
    \caption{Hypothesis $h_3$ revised by Frattini et al.~\cite{frattini2024applying}}
    \label{fig:rq:h3}
\end{figure}

Second, the authors produced another piece of evidence $e_4$ that both revises the causal assumptions of $e_2$ (i.e., replaced $h_2$ with $h_3$) and performs a Bayesian data analysis $m_2$ on the new data $d_2$.
The revised hypothesis $h_3$, visualized in \Cref{fig:rq:h3}, made several changes:

\begin{enumerate}
    \item Missing actors and domain objects behave the same and, hence, can be aggregated to the number of missing \textit{entities} in the domain model.
    \item The number of missing associations may further be impacted by the education, task experience, and domain knowledge of a participant.
    \item The amount of time that a participant took to generate the domain model may influence its completeness.
\end{enumerate}

Evidence $e_4=E(h_3, d_2, m_2)$ concludes that the use of passive voice in NL requirements leads to more missing associations in about 45\%, to fewer in 25\%, and to an equal amount in 30\% of all cases on average.
$e_4$ agrees with $e_2$ in that the impact of passive voice on the number of missing associations from domain models is not strictly negative, as the likelihood of missing an association remains below 50\%.
However, $e_4$ suggests that the negative impact is more likely than assumed by $e_2$ (45\% instead of 34\%).
Since both pieces of evidence were produced under different causal assumptions ($h_2$ in $e_2$ and $h_3$ in $e_4$), these numbers are not directly comparable.
Again, the authors of this follow-up study did not validate whether $h_3$ was more valid than $h_2$.

\begin{highlightbox}{Insight 1}
    The research synthesis produced a variance theory suggesting that the use of passive voice in NL requirements has a moderate effect on the number of associations missing from domain models.
    However, the application of the framework reveals that several transitions (i.e., replacing hypotheses) were not validated.
    Hence, their contribution to the validity of the variance theory remains questionable.
\end{highlightbox}

\subsection{In-depth Research Synthesis}
\label{sec:application:critique}

The four pieces of evidence $e_1$-$e_4$ from the three studies~\cite{femmer2014impact,frattini2024second,frattini2024applying} were produced without an explicit framework for managing variance theories.
This caused several steps in the evolution of the variance theory about the impact of passive voice on domain modeling to be of unclear validity.
Evidence $e_2$ conflates a revision with a reanalysis, i.e., it replaces both the hypothesis ($h_2$ for $h_1$) and the analysis method ($m_2$ for $m_1$).
The differing conclusions obtained from $e_2$ can, therefore, not be traced clearly to either of these two changes.

In the following in-depth analysis, we will zoom in on the step of evolution between the original evidence $e_1$ and the evidence from the first follow-up study $e_2$.
We disentangle the sub-steps according to the proposed framework, which allows us to retrospectively assess the evolution performed by Frattini et al.~\cite{frattini2024second} but also guide future contributions to this variance theory.
\Cref{tab:application:quality:detail} visualizes the deconstructed pieces of evidence between $e_1$ and $e_2$.
The rows between $e_1$ and $e_2$ show the disentangled sub-steps between the two pieces of evidence.
The rows after $e_2$ show additional sub-steps that would have been the more valid path to pursue had the authors of the follow-up study~\cite{frattini2024second} disentangled the sub-steps.

\begin{table*}[!ht]
    \centering
    \caption{Decomposed steps between $e_1$ and $e_2$}
    \label{tab:application:quality:detail}
    \begin{tabular}[b]{llllll}
        \toprule
        \textbf{VCG} & \textbf{Evolution Type} & \textbf{Ref.} & \textbf{Hyp.} & \textbf{Analysis Method} & \textbf{Conclusion} \\
        \midrule
        \multirow[c]{8}{*}{ \includegraphics[height=72.5px]{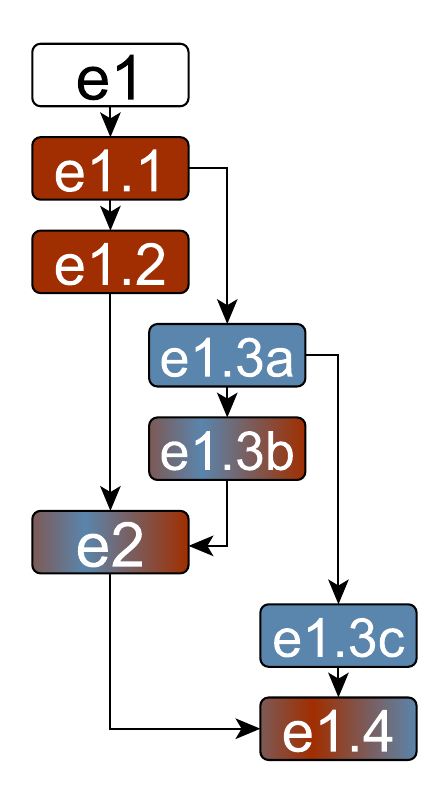}} & Original Study & \cite{femmer2014impact} & $h_1$ & $m_1$: Mann-Whitney U test & $p=0.001$ \\
        \hline
        & Reanalysis & & $h_1$ & $m_{1.1}$: linear model & $ci=[0.23, 0.84]$ \\
        & Reanalysis & & $h_1$ & $m_2$: Bayesian model & $[-0.15, \sim0.34, +0.51]$\\
        & Revision & & $h_{2a}$ & $m_{1.1}$: linear model & $ci=[0.17, 0.77]$ \\
        & Revision & & $h_{2}$ & $m_{1.2}$: linear mixed model & $ci=[-0.17, 0.82] $\\
        \hline
        & Revision/Reanalysis & \cite{frattini2024second} & $h_2$ & $m_2$: Bayesian model & $[-0.17, \sim0.49, +0.34]$ \\
        \hline
        & Revision & & $h_{2c}$ & $m_{1.2}$: linear mixed model & $ci=[0.03, 0.92]$ \\
        & Reanalysis/Revision & & $h_{2c}$ & $m_2$: Bayesian model & $[-0.14, \sim0.47, +0.39]$ \\
        \bottomrule
    \end{tabular}
  \end{table*}

\subsubsection{Reanalysis $e_{1.1}$}

The frequentist NHST $m_1$ is not directly comparable to a Bayesian data analysis $m_2$.
Hence, an intermediate step is necessary.
A simple reanalysis is to replace the Mann Whitney U test $m_1$ with a linear regression model $m_{1.1}$.
At the core, a linear model that regresses the rank-transformed outcome variable on a single predictor is equivalent to the Mann Whitney U test~\cite{lindelov2019common}.
Fitting a linear model $Asc^- \sim passive$ to the data $d_1$ produces a coefficient of $\beta_{passive}=0.53$ and a 95\% confidence interval of $ci_{e_{1.1}}(passive) = [0.23, 0.84]$.
The confidence interval is not consistent with 0, i.e., it does not contain 0.
Therefore, the conclusion of $e_{1.1}$ agrees with the conclusion of $e_1$ in its suggestion that the use of passive voice has a statistically significant impact on the number of missing associations from domain models.

\subsubsection{Reanalysis $e_{1.2}$}

Replacing the linear regression model $m_{1.1}$ with a Bayesian data analysis $m_2$ produces evidence $e_{1.2}$, a strict reanalysis of $e_1$ and $e_{1.1}$ as it only changes the analysis method but retains hypothesis $h_1$ and data $d_1$.
The statistical comparison between $m_2$ and $m_{1.1}$ relies on existing literature that explains, at length, how Bayesian methods have a higher conclusion validity.
They preserve uncertainty~\cite{torkar2020bayesian}, do not make use of the invalid probabilistic extension of the modus tollens~\cite{furia2019bayesian}, and allow modeling the response variable with other distributions than the normal distribution~\cite{mcelreath2018statistical,gren2021possible}.
In the case of $e_{1.2}$, the response variable can be modeled with a binomial distribution $Asc^- \sim B(n, p)$ where $n$ represents the number of expected associations and $p$ the likelihood of missing one association.
This distribution encodes the ontological assumption that the number of potentially missing associations is bounded by the number of expected associations in the gold standard of the expected domain model, i.e., a participant in the experiment cannot miss more associations than the gold standard contained.
This assumption could not be implemented in the frequentist analysis methods that assumed the (rank-transformed) outcome variable to be normal.
Therefore, $m_2$ exceeds $m_{1.1}$ in conclusion validity.
Evidence $e_{1.2}$ concludes that the use of passive voice leads to more missing associations in 51\%, fewer in 15\%, and an equal amount in 34\% of all cases.
This conclusion still supports that the use of passive voice has an impact on the number of missing associations, but remains more cautious.

\subsubsection{Revision $e_{1.3a}$}

\begin{figure}
    \centering
    \includegraphics[width=\linewidth]{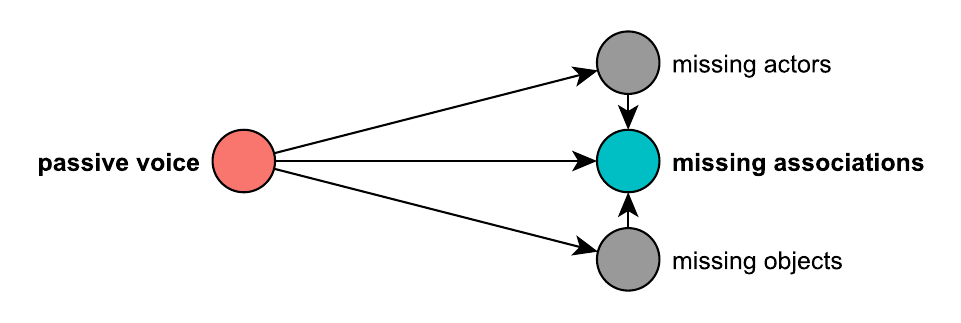}
    \caption{Hypothesis $h_{2a}$ with the purpose of debiasing}
    \label{fig:rq:h2a}
\end{figure}

Since $e_{1.1}$ replaces the simple NHST with a linear model, we can systematically revise the hypothesis $h_1$ by adding additional predictors.
However, the revision of $h_1$ to $h_2$ in $e_2$ performed two separate revisions with different purposes.
Firstly, the authors added assumed causal relationships of the number of missing actors and domain objects on the number of missing associations (additional assumption 1 in \Cref{sec:application:evolution:wsese}).
\Cref{fig:rq:h2a} visualizes the hypothesis $h_{2a}$ resulting from adding just this assumption to $h_1$.
$h_{2a}$ is a sub-graph of the previously introduced $h_2$ (\Cref{fig:rq:h2}), only missing the additional variables.

To determine which of the two hypotheses $h_1$ and $h_{2a}$ has greater internal validity, we need to assess the testable implications of the models.
In particular, $h_1$ implies that the outcome variable $Asc^-$ is independent of the number of missing actors $Act^-$ and domain objects $Obj^-$ when conditioning on the use of passive voice.
Comparing the coefficients of the two fit linear models $e_{1.1}$ and $e_{1.3a}$ shows that $\beta_{passive}$ only shifts slightly but remains inconsistent with 0.
On the other hand, $\beta_{Obj^-}$ is also inconsistent with 0, confirming that the outcome variable is not independent of the mediator $Obj^-$.
This warrants their inclusion in the hypothesis and confirms that $h_{2a}$ is more internally valid than $h_1$.

\subsubsection{Revision and Reanalysis $e_{1.3b}$}

Secondly, the authors of the follow-up study~\cite{frattini2024second} performed a revision with the purpose of increasing the precision of the estimate. 
To this end, they included additional variables that were recorded in $d_1$ to the hypothesis, resulting in $h_2$ as visualized in \Cref{fig:rq:h2}.
Two of the newly included predictors---the participants' skill and the requirements' complexity---are modeled as random effects, which requires extending the linear model $m_{1.1}$ to a linear mixed model $m_{1.2}$.
Consequently, this step constitutes both a revision and a reanalysis.
Both of these evolutions need to be assessed individually.

To determine which of the two analysis methods has a greater conclusion validity, we can assess the statistical properties of the pieces of evidence.
For instance, the residuals of a linear model should be independent and identically distributed (iid)~\cite{west2022linear}.
This property can be determined graphically by inspecting a histogram of the residuals, a QQ-plot, by running tests like the Durbin-Watson test, or other diagnostics.
Both graphical analyses (to be found in our replication package~\cite{tse2024replication}) and the statistical tests suggests that the residuals of $e_{1.1}$ are not iid:
The Shapiro-Wilk test suggests a significant deviation from a normal distribution of the residuals ($p=3.26e-05$), the Durbin-Watson test suggests an autocorrelation greater than 0 ($p=0.07$), only the Breusch-Pagan test does not suggest heteroscedasticity ($p=0.11$).
Consequently, applying a linear model $m_{1.1}$ may lead to invalid conclusions, and the conclusion validity of a linear mixed model $m_{1.2}$ is greater~\cite{west2022linear}.

To determine which of the two hypotheses has the greater internal validity, we evaluate their predictive power.
Since the analysis methods differ, conventional metrics like $R^2$ and its variants are ineligible, as they apply only to one of the two methods.
Instead, we calculate the AIC, which applies to both~\cite{vaida2005conditional}.
The two pieces of evidence achieve scores of $AIC(e_{1.3a})=249.1$ and $AIC(e_{1.3b})=251.2$.
The score differential of about 2 points is considered negligible when interpreting the AIC values~\cite{greenwood2021intermediate}.
Hence, there is no strong evidence that $h_2$ is more internally valid than $h_{2a}$.
However, the confidence interval of $\beta_{passive}$ concluded by $e_{1.3b}$ is $ci_{e_{1.3b}}(passive)=[-0.17, 0.82]$.
$e_{1.3b}$ is the first piece of evidence suggesting that the use of passive voice does not have a statistically significant impact on the number of missing associations in domain models.

\subsubsection{Revision/Reanalysis $e_2$}

This leads to the target evidence $e_2$, which can now be considered a strict revision of $e_{1.2}$ and a strict reanalysis of $e_{1.3b}$.
As such, the eligibility of these evolution steps can be assessed individually.
The reanalysis of $e_{1.3b}$ to $e_2$ again relies on literature explaining the advantage of Bayesian over frequentist methods~\cite{furia2019bayesian,torkar2020bayesian,mcelreath2018statistical,gren2021possible}.
The revision of $e_{1.2}$ to $e_2$ again requires two steps and needs to determine that the inclusion of assumptions both de-biases the estimate and increases its precision.

The conclusion of $e_2$ is, as presented in \Cref{sec:application:evolution:wsese}, that using passive voice is less impactful than originally assumed~\cite{femmer2014impact}.
The decomposed sub-steps make this more evident, as $e_2$ claims that passive voice causes more missing associations in only 34\% of all cases, other than $e_{1.2}$, which claimed it to be 51\%.
However, the target evidence $e_2$ is subject to several shortcomings due to the conflation of the revision with the reanalysis.
Firstly, the follow-up study~\cite{frattini2024second} attributes the differing conclusion mainly to the use of Bayesian methods (i.e., the reanalysis) rather than the changed hypothesis (i.e., the revision).
The detailed analysis, though, clearly shows that only $e_{1.3b}$, i.e., the revision to $h_2$, made the conclusion more cautious.
Secondly, the follow-up study never assesses whether $h_2$ has greater internal validity than $h_1$ and deserves to be subscribed to.
The model comparison in scope of the revision $e_{1.3a}$ shows that the support for $h_2a$ over $h_1$ is actually minimal and, hence, more debatable than the follow-up study makes it seem.
Adherence to the proposed framework for managing variance theories revealed these shortcomings and made the reliability of each step transparent.

\subsubsection{Revision $e_{1.3c}$}

\begin{figure}
    \centering
    \includegraphics[width=\linewidth]{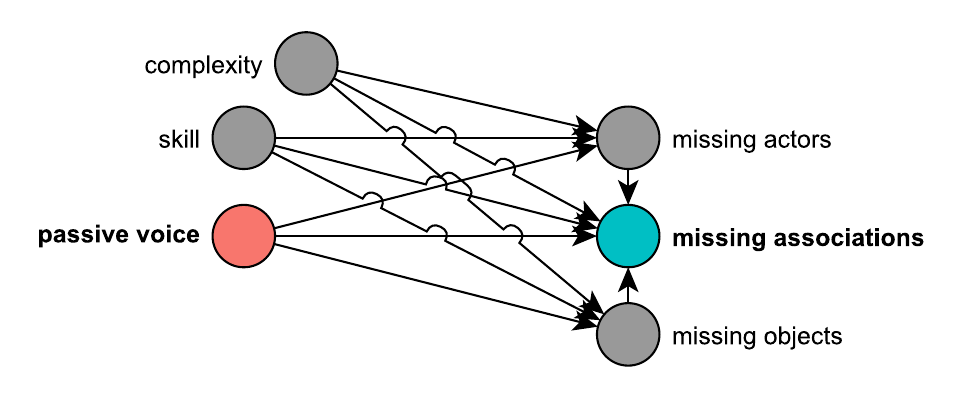}
    \caption{Hypothesis $h_{2c}$}
    \label{fig:rq:h2c}
\end{figure}

The stepwise evolution according to the proposed framework revealed that $h_2$ has little support over $h_{2a}$ as the AIC scores are close enough together to consider both hypotheses of equal predictive power~\cite{greenwood2021intermediate}.
However, the diagnostics of $e_{1.3b}$ reveal that the inclusion of random effects had a strong impact:
While $e_{1.3b}$ only has a marginal $R^2$ value of 0.184, it has a conditional $R^2$ value of 0.561.
This indicates that the random effects explain a lot more of the variance of the outcome variable than the fixed effects~\cite{nakagawa2013general}.
The benefit of the random effects in $h_2$ may be offset by the number of predictors, as the AIC metric penalizes an increased number of predictors to avoid overfitting~\cite{greenwood2021intermediate}.
Hence, the authors could have formulated the competing hypothesis $h_{2c}$ (shown in \Cref{fig:rq:h2c}), which retains the provenly effective random effects for requirements' complexity and participants' skill but discards the fixed effects of academic and industrial experience.
The operationalization of both of these fixed effects is questionable, as they were simply measured on an ordinal scale with four levels.
Because the construct validity of this operationalization is questionable, the inclusion of these fixed effects might not benefit the estimation and rather overfit the estimation.

For model comparison to assess the predictive power of the new piece of evidence, we can use the Akaike information criterion.
The resulting $AIC(e_{1.3c}) = 241.4$ is significantly---i.e., more than 2 units~\cite{greenwood2021intermediate}---lower than $AIC(e_{1.3a})=249.1$ and $AIC(e_{1.3b})=251.2$.
Consequently, $h_{2c}$ shows the greatest internal validity and should have been used subsequently instead of $h_2$.
Evidence $e_{1.3c}$ concludes that passive voice has an impact of $ci_{e_{1.3c}}(passive)=[0.03, 0.92]$, which is still not consistent with 0 but broader than $ci_{e_{1.3b}}(passive)=[0.17, 0.77]$.
This suggests that passive voice does have an impact on the number of missing associations, though the strength of the impact varies more.

\subsubsection{Reanalysis/Revision $e_{1.4}$}

The more rigorous target evidence of the follow-up study~\cite{frattini2024second} would have been $e_{1.4}$, which applies the analysis method with the highest conclusion validity---Bayesian data analysis $m_2$---to the data $d_1$ under the hypothesis with the highest internal validity---hypothesis $h_{2c}$, not $h_2$.
This piece of evidence classifies as a reanalysis of $e_{1.3c}$ (as it only substitutes $m_{1.2}$ with $m_2$) and a revision of $e_2$ (as it substitutes $h_2$ with $h_{2c}$).
The validity of the reanalysis is again based on previous literature~\cite{furia2019bayesian,torkar2020bayesian,mcelreath2018statistical,gren2021possible} and the revision on the model comparison regarding predictive power.
In the case of comparing two Bayesian models, we can use the leave-one-out (LOO) cross-validation~\cite{vehtari2017practical}.
The LOO comparison favors $e_{1.4}$ over $e_2$ as expected based on the previous comparison of frequentist models using the AIC metric.
This piece of evidence $e_{1.4}$ concludes that the use of passive voice leads to more missing associations in 39\%, fewer in 14\%, and an equal amount in 47\% of all cases.
Considering this the most valid piece of evidence about the impact of passive voice at the time of the follow-up study~\cite{frattini2024second}, the variance theory would suggest that passive voice does have an impact on the number of missing associations from domain models, though not strictly and only in about 40\% of all cases.

\begin{highlightbox}{Insight 2}
    The application of the framework to disentangle the omitted sub-steps between $e_1$ and $e_2$ revealed that $h_2$ was not the optimal improvement over $h_1$, but instead, $h_{2c}$ would have been.
    Additionally, the random effects modeling participants' skill improved the precision of the ACE estimation far greater than variables like experience (measured on an ordinal scale).
    This indicates that years of experience is an insufficient operationalization to represent the latent context variable of modeling skill.
    Finally, the adjusted conclusions can be attributed more to the revision than to the reanalysis.
\end{highlightbox}

\section{Discussion}
\label{sec:discussion}

The proposed framework enables researchers to systematically manage variance theories.
The framework offers a definition of empirical, quantitative evidence, a clear terminology of the relationship between two pieces of evidence, and guidance on how to determine which piece of evidence has the greater validity.
The discussion at the ISERN workshop agreed on three implications for research practice (\Cref{sec:discussion:implications}) while also acknowledging several limitations (\Cref{sec:discussion:limitations}) that necessitate future work (\Cref{sec:discussion:future}).

\subsection{Implications}
\label{sec:discussion:implications}

First, our terminology of evolution types allows primary studies to clearly position themselves in relation to the existing body of knowledge.
Authors contributing a replication, revision, or reanalysis of a phenomenon with at least one prior study, can label their follow-up study with the respective evolution type to indicate how they advance the scientific frontier of an existing body of knowledge.
Additionally, the framework specifies how authors can determine whether their follow-up study is of greater validity or not, e.g., via model comparison in the case of a revision for de-confounding.
We hope this helps researchers to publish also negative results, as these indicate possible ``dead ends'' in a research topic.

Second, the proposed framework allows researchers to relate pieces of evidence in secondary studies more clearly.
We hope to inspire more rigorous literature reviews about phenomena that synthesize empirical, quantitative evidence to variance theories.
Visualizing the evolution of a variance theory in the form of a version control graph as demonstrated in \Cref{sec:application:evolution,sec:application:critique} helps to communicate the progress within a research field.

Finally, the application of the proposed framework prospectively shapes future research.
With the scientific frontier of a research field determined and analytically supported, an application of the framework can inform the design of future studies.
For example, the hypothesis with the highest internal validity informs new research about the factors that need to be measured.
Similarly, the analysis method with the highest conclusion validity informs how to perform the data analysis on collected data.
This way, researchers can coordinate research agendas working towards a shared variance theory about a quantitative phenomenon.

\subsection{Limitations}
\label{sec:discussion:limitations}

We differentiate the limitations of the framework itself from the limitations to the adoption of the framework in SE research.
One significant limitation of the framework is that it depends on reliable operationalizations of the concepts involved in the phenomena under study.
The dimension of validity that the proposed framework does not systematically address is construct validity. 
A natural option would have been to define hypotheses on the concept, not on the indicator level.
For example, the phenomenon discussed in \Cref{sec:application} could have been abstracted to the impact of requirements quality on domain modeling performance instead of passive voice on the number of missing associations.
This would extend the types of evolution in \Cref{fig:framework} by one that challenges the operationalization of concepts, i.e., that improves the construct validity of the measurements.
In the example, this would mean challenging whether passive voice is a valid indicator of requirements quality or whether the number of missing associations adequately represents domain modeling performance.
However, we opted against this as we could not find a consensus on the systematic comparison of construct validity in SE research.
Instead, we assume all variables involved in hypotheses to be on an indicator level and delegate their abstraction to concepts outside of the framework.
Should a competing piece of evidence aim to improve the construct validity and propose a hypothesis in which at least one concept is operationalized differently, then these new pieces of evidence are incommensurable.

Additionally, the proposed framework depends on the rigor of the applied research methods.
SE research has been shown to be subject to researcher bias~\cite{shepperd2014researcher,romano2021researcher}, which implies that the conclusions drawn from evidence differ not only depending on the hypotheses, data, and methods involved but also based on the people that produced the evidence.
This necessitates manually ensuring a sufficient level of rigor in pieces of evidence that shall be included in a body of knowledge.

Furthermore, the usefulness of the proposed framework depends on researchers' adoption of it.
One limitation to this is the complexity of the framework itself. 
While methodologies for replications are well established in SE research~\cite{baldassarre2014replication,gomez2010replications,shepperd2013combining}, approaches for systematic comparison of causal hypotheses via model comparison are not yet commonplace~\cite{siebert2023applications} and the selection of analysis methods often follows conventions.
The framework offers a principled approach that places revisions and reanalyses into a relationship with the existing body of knowledge but requires that scholars familiarize themselves with model comparison and challenge statistical conventions.
We envision that the proposed framework will initially be most useful to researchers or research groups willing to immerse themselves in these methods by putting their own pieces of evidence about a shared phenomenon into relation.
Later, we hope that larger applications like systematic literature reviews or adoption by a whole community will become possible.

\subsection{Future Work}
\label{sec:discussion:future}

The primary goal of our future work will be to communicate this framework and offer support in adopting it.
This means not just presenting the framework as is but connecting the elements of the framework as shown in \Cref{fig:framework} with literature that helps scholars to apply the presented relationships.
For example, the replication branch can be supported with definitions~\cite{baldassarre2014replication,gomez2010replications}, philosophical stances~\cite{nosek2020replication}, and advice on performing meta-analyses~\cite{shepperd2013combining}.
We envision that this communication support will take the form of a web platform that makes both the framework and the recommended resources, like further reading and demonstrations, accessible.

The second goal of future work is to further support the application of the framework by evolving the aforementioned platform under the umbrella of the ISERN community.
We aim to extend the platform into a system where pieces of evidence can be submitted and that visualizes the evolution of a body of knowledge about a phenomenon.
This feature of the platform will resemble version control systems like git for empirical evidence.
Such a platform will host the current body of knowledge about SE phenomena. 
It will support researchers both in finding the most recent contributions to a phenomenon and inform future study designs by indicating the hypotheses with the highest validity.
This platform will address the final shortcoming mentioned in \Cref{sec:related:shortcomings} and replace static, retrospective research synthesis with a dynamic, continuous process.
Approaches to automate model comparison and meta-analysis may pave the way towards an even more dynamic process.
Additionally, such a platform may support the initially mentioned knowledge translation by offering an interface to practitioners to obtain synthesized---and, therefore, more valid---conclusions from variance theories.
These quantitative conclusions serve as decision support, for example when determining whether to adopt TDD practices~\cite{rafique2012effects} or whether to remove passive voice from requirements documents~\cite{frattini2024applying}.

Finally, we aim to demonstrate the application of the framework to additional fields of SE research.
The field of requirements quality was chosen due to the authors' familiarity with it and since it contains coherent yet manageable pieces of evidence.
Fields that aim to produce variance theories of phenomena, like the impact of TDD on code quality and developer effectiveness~\cite{rafique2012effects}, are eligible for such an application.
We envision a special type of literature review emerging from the framework, focusing on quantitative primary studies about a particular phenomenon.

\section{Conclusion}
\label{sec:conclusion}

To effectively progress the development of variance theories from quantitative, empirical evidence, SE research needs a research synthesis approach that extends beyond the meta-analysis of homogeneous replications.
In this article, we define quantitative, empirical evidence, propose a typology of relationships between two pieces of evidence, and offer guidance on determining which piece of evidence has greater validity.
Expressing research agendas through this framework enables the systematic management of variance theories and guides SE research toward producing more rigorous conclusions.

\section*{Acknowledgments}
This work was supported by the KKS foundation through the S.E.R.T. Research Profile project and the GIST project at Blekinge Institute of Technology.

\bibliographystyle{IEEEtran}
\bibliography{references}

\end{document}